\documentclass[12pt]{article}
\pdfoutput=1
\usepackage{natbib}
 \bibpunct{(}{)}{;}{a}{,}{,}
\usepackage{graphicx}
\usepackage{amsmath,amssymb,amsthm}
\usepackage{bm}
\usepackage{rotating}
\usepackage[lined,ruled]{algorithm2e}
\usepackage{multirow}
\usepackage{rotating}
\usepackage{multicol}
\usepackage{titlesec}
\usepackage{latexsym}
\usepackage[pdftex,colorlinks=true,linkcolor=blue,citecolor=blue,urlcolor=blue,bookmarks=false,pdfpagemode=None]{hyperref}
\usepackage{url}
\usepackage{verbatim}
\usepackage{fancyhdr}
\usepackage{setspace}
\usepackage{paralist}
\usepackage{lineno}
\usepackage[top=1in, bottom=1in, left=1in, right=1in]{geometry}
\usepackage{lscape}
\usepackage{dcolumn}

\newcolumntype{.}{D{.}{.}{-1}}

\usepackage{enumerate} 
\newtheorem{theorem}{Theorem}[section]

\newtheorem{assumption}{Assumption}

\newtheorem{definition}{Definition}

\begin{document}
\pagestyle{empty}


\title{Randomization and The Pernicious Effects of Limited Budgets on Auction Experiments}

\author{
 Guillaume W. Basse$^\dagger$ \and Hossein Azari Soufiani$^\ddagger$, \and Diane Lambert
 \thanks{
Guillaume Basse is a graduate student in the Department of Statistics at 
Harvard University (\href{mailto:gbasse@fas.harvard.edu}{gbasse@fas.harvard.edu})
 and his work was supported by a Google Fellowship in Statistics for 
North America. Hossein Azari Soufiani is a research scientist at Google Research
 (\href{mailto:azari@google.com}{azari@google.com}). Diane Lambert is a 
 retired research scientist  (\href{mailto:exdlambert@gmail.com}{exdlambert@gmail.com}).  
 This work was completed at Google Research. 
The authors thank Google's Max Lin and George Levitte for their 
help, as well as Google NYC's statistics group for early feedback.
We also thank three anonymous reviewers for their constructive 
feedback.
}}

\date{}

\maketitle
\thispagestyle{empty}

\newpage
\begin{abstract}
  Buyers (e.g., advertisers) often have limited financial and
  processing resources, and so their participation in auctions is
  throttled. Changes to auctions may affect bids or throttling and any
  change may affect what winners pay. This paper shows that if an A/B
  experiment affects only bids, then the observed treatment effect is
  unbiased when all the bidders in an auction are randomly assigned to
  A or B but it can be severely biased otherwise, even in the absence
  of throttling. Experiments that affect throttling algorithms can
  also be badly biased, but the bias can be substantially reduced if the budget
  for each advertiser in the experiment is allocated to separate
  pots for the A and B arms of the experiment.

\vfill
\end{abstract}

\newpage
\singlespacing
\small
\tableofcontents
\normalsize
\singlespacing

\newpage
\pagestyle{plain}
\setcounter{page}{1}


\section{INTRODUCTION}
A search query generates a request for ads to show with search
results. A user's visit to a webpage generates a request to an ad
exchange for page ads. In either case, advertisers are chosen to
participate in an auction that chooses the ad to be shown. Advertisers
often cannot pay for or process all auction requests they are
eligible for, so the requests passed on to them are throttled to meet
their quota constraints.

Auction parameters like reserve prices and throttling schemes can
affect an advertiser's payments and an exchange's revenue, so A/B
experiments are run to test ideas for improving outcomes for
advertisers and the ad exchange. \cite{lucking1999using}
and \cite{einav2011learning} experimented with auction formats,
\cite{reiley2006field} and \cite{ostrovsky2011reserve} experimented
with reserve prices, and \cite{ausubel2013experiment} experimented
with budget constraints.

Usually, A/B experiments are analyzed by assuming that the two
experiment arms, traditionally called treatment and control, are
independent. But, as \cite{blake2014marketplace} explains,
independence fails when the demands of the treatment and control arms
affect each other. Such {\em interference} is unavoidable when some
advertisers in an auction are in treatment and some in
control. \cite{kohavi2009controlled} recognizes that some
randomization schemes can give misleading treatment estimates for
auction experiments. For more insight into interference in other
applications see \cite{halloran1995causal}, \cite{hudgens2012toward},
\cite{rosenbaum2007interference}, \cite{tchetgen2010causal},
\cite{aronow2013estimating}, \cite{eckles2014design}, \cite{athey2015exact} and
\cite{airoldi2012}. Optimal experiment design in the presence of
interference has been explored by \cite{david1996designs}, 
\cite{eckles2014design} and \cite{walker2014design}.

This paper uses the framework of causal models and interference to
shed light on auction experiments. Section 2 introduces the main
elements of our model:
\begin{enumerate}
\item potential outcomes that describe what each advertiser would bid
  if assigned to treatment or if assigned to control,
\item {throttling} algorithms that determine which of the advertisers
  {\em eligible} to respond to a request for an ad are called to its auction,
\item {\em bid treatments} that affect what advertisers bid and {\em
    throttling treatments} that affect when they are called to bid,
  and
\item effects on the total daily payment of an advertiser or the
total daily revenue of the ad exchange.
\end{enumerate}
Section 3 defines two randomization schemes for auction
experiments. In {\em query randomization} each request for an ad is
randomized to treatment or control, so all participants in an auction
are in treatment or all are in control. In {\em (query, advertiser)
  randomization}, each advertiser that is eligible for a query is independently
assigned to treatment or control, so treatment bidders and control bidders
can compete in the same auction. With either kind
of randomization, an advertiser can be assigned to treatment for some
queries and to control for others.  (Query, advertiser) randomization is
undesirable because it introduces interference, but it may be
unavoidable if only some advertisers are included in an experiment,
perhaps to avoid revealing a possible change in auction algorithms
before it has proven useful. The remainder of the paper explores bias
and variance of estimated effects for bid and throttling treatments
under query and (query, advertiser) randomization.

To establish the ideas, Section 4 shows what can go wrong when
treatment and control bidders compete in the same auction. Section 5
introduces budget (processing or financial) throttling. In {\em Split
  quota} experiments each advertiser has separate budgets for
treatment and control queries. In {\em joint quota} experiments the
advertiser draws against the same budget for all its
queries. Simulations in Section 6 suggest that estimated treatment
effects can be severely biased under (query, advertiser) randomization
regardless of whether the budget is split or joint for both bid and
throttling experiments. Estimated treatment effects for throttling
experiments are biased for both query and (query, advertiser) randomization,
but the bias is much smaller for split quota than for joint quota
experiments.

\section{CAUSAL MODELS FOR AUCTIONS}
This section casts auction experiments as causal models in which each
advertiser has two potential bids for each query: the bid that would
be made if the advertiser were assigned to treatment for that query
and the bid if assigned to control. Of course, only one potential bid
can be observed. This section points out further subtleties that
result from advertiser competition and quota throttling.

\subsection{Potential Bids}
To start, suppose there is no throttling, so advertisers bid in all
auctions for which they are eligible.  The raw data for a sample of
auctions is then a set of vectors $(q,a,B,Y)$, where $q$ denotes a
query, $a$ an advertiser, $B$ the advertiser's bid, and $Y$ the
advertiser's payment. We consider only auctions like first and second
price auctions in which the payment is positive if the advertiser wins
the auction and zero otherwise. Define $N_q$ to be the number of
unique queries and $N$ the total number of (query, advertiser) pairs,
typically considered over the course of one day.

In an experiment, a (query, advertiser) pair is assigned to either
treatment or control. Let $Z = (Z_1, \ldots, Z_N)$ be the treatment
assignment vector, where $Z_i = 1$ if (query, advertiser) pair $i$ is
assigned to treatment, and $Z_i = 0$ if assigned to control. Following
the potential outcomes framework of \cite{rubin1990po}, each eligible
advertiser for a query has a bid and payment that will be observed if the pair is
assigned to control and a possibly different bid and payment that will
be observed if assigned to treatment. That is, the {\em potential bids} for the $N$
(query, advertiser) pairs are $B(Z) = (B_1(Z), \ldots, B_N(Z)) $ and the
potential payments are $Y(Z) = (Y_1(Z), \ldots, Y_N(Z))$.

An advertiser does not know which other advertisers are eligible for a
query so its bid is independent of all other bids for the query, which
implies that the following assumption given in \cite{rubin1980comment}
holds.
\begin{assumption}[Stable Unit Treatment Value (SUTVA)]
There is only one
version of the treatment and there is no interference in the potential
bids for eligible advertisers in treatment and control.
\end{assumption}\label{sutva}
Because SUTVA holds for bids, the potential bids for all query,
eligible advertiser pairs under treatment $B(\vec{1})$ and under control
$B(\vec{0})$ can be considered separately.

\subsection{Potential payments}
If bidder i's payment is positive, then any other
bidder in the auction pays zero, so a treatment that affects what
advertisers bid can affect the payment for both treatment and control
advertisers. For example, suppose there are three advertisers in an auction
with the
potential bids given in the following table and the winner pays what it bid.

\begin{table}[h!]
	\centering
	\begin{tabular}{|c|c|c|}
		\hline
		advertiser & B(1) & B(0) \\
		\hline
		$b_1$ & 5 & 2 \\
		\hline
		$b_2$ & 4 & 4 \\
		\hline
		$b_3$ & 3 & 1 \\
		\hline
	\end{tabular}
\end{table}

If all three advertisers participate in the auction and advertiser 2 is the
only one assigned to treatment, then it wins and pays
$Y_2( (0,1,0) ) = 4$. However, if advertiser 1 is also assigned to treatment and
participates in the auction then advertiser 2 loses and pays nothing.
This is an example of interference: the payment of an advertiser is
affected by the assignment of other advertisers in
the auction to treatment and control. Hence, SUTVA does not hold for payments.

We make the following assumption about payments.
\begin{assumption} The potential payment of an advertiser eligible for a query
depends only on its assignment to treatment or control and the
assignments of the other eligible advertisers for the query. That is
\begin{equation}
	Y_i(Z) = Y_i(Z_{q[i]})
\end{equation}
where $Z_{q[i]}$ is the subset of the assignment vector $Z$ on (query,
advertiser) pairs for which the query is $q[i]$.
\end{assumption}
We extend this assumption to auctions with quota throttling in
Section 2.3.

The interference structure allowed by Assumption 2 is similar to that
given in \cite{rosenbaum2007interference} and \cite{hudgens2012toward}
and can be seen as a special type of \textit{effective treatment}
(\cite{manski2013identification}) or \textit{exposure mapping}
(\cite{aronow2013estimating, eckles2014design}). However, those papers
do not consider quota throttling.

\subsection{Quota throttling}
\label{section:quota-intro}
Advertisers are generally {\em quota constrained}, meaning that they
have insufficient budget or infrastructure to process all the queries
they could be sent (\cite{chakraborty2010selective}). In that case,
some queries are dropped or {\em throttled} to meet the quota
constraints. That is, there is a vector
$W(Z) = (W_1(Z), \ldots, W_N(Z))$ with $W_i(Z) = 0$ if (query, advertiser) 
pair $i$ is dropped and $W_i(Z) = 1$
otherwise. Note that throttling can depend on the assignment $Z$ to
treatment and control. We assume random dropping according to a {\em
  throttling distribution} $p(W(Z)|Z)$. In the presence of throttling, Assumption~2 
  means that the potential payment of an advertiser depends only on its
  assignment to treatment or control, and on the assignment of the advertisers
  for the same query who were not throttled. Henceforth, advertisers who 
  participate (bid) in auctions are called bidders.

Because how much a bidder in an auction pays depends on the other
bidders in the auction, the potential payment for any
advertiser $a$ is random when at least one other advertiser for the
query is quota constrained, even if advertiser $a$ is unconstrained.

\subsection{Bid treatments and throttling treatments}
Loosely speaking, an experiment about bids is designed to test
whether a change to bidding rules, such as a change to the reserve or
``floor'' price for auctions, matters when throttling rules are
unchanged. (See \cite{reiley2006field} and \cite{ostrovsky2011reserve} for
examples.)
\begin{definition}[Bid Treatment]
	Under a {\em bid treatment}, for all $Z$, the throttling
        distribution $W$ satisfies
	\begin{equation}
		W(Z) | Z \sim (W(Z) | Z=\vec{1}) \sim (W(Z) | Z=\vec{0}).
	\end{equation}
\end{definition}
In words, a bid treatment affects only potential bids, so a given
(query, advertiser) for an eligible advertiser has the same probability of
being dropped regardless of how other eligible bidders are assigned
to treatment and control. That condition holds if queries are throttled before
the advertiser's bid is known.

An experiment about throttling is designed to understand
whether changing the rules for meeting quota constraints matters
if bidding parameters like reserve prices are unchanged. (See the
selective callout algorithm in \cite{chakraborty2010selective}.) 
\begin{definition}[Throttling Treatment]
	Under a {\em throttling treatment}, for all $Z$, the potential
        bids satisfy
	\begin{equation}
		B(Z) = B(\vec{0}) = B(\vec{1}).
	\end{equation}
\end{definition}
In a throttling experiment, any eligible advertiser would bid the same amount,
whether it is throttled or not. This is true if advertisers do not reveal their
bids before throttling.

\subsection{Effects on revenue}

If no advertiser is quota constrained, then no advertiser is dropped from a
query and the effect $\tau$
of the treatment on the total revenue of the ad exchange compares the
revenue when every (query, advertiser) pair is treated to the
revenue when every (query, advertiser) pair is in control:
\begin{equation}
	\tau = \sum_i^N (Y_i(\vec{1}) - Y_i(\vec{0}))
\end{equation}
where $Y_i$ is the payment of (query, advertiser) pair $i$.  If there
are quota constraints, then $\tau$ is affected by random dropping so
the effect on total revenue is $\tau^* = E(\tau)$, taking the
expectation under the dropping scheme $p(W)$. Note that the sum in
this case is taken over the $N$ (query, eligible advertiser) pairs.

The effect $\tau_a$ of treatment on the revenue generated
by a given advertiser $a$ when there are no quota constraints is given by
\begin{equation}
	\tau_a = \sum_{i:a[i] = a} (Y_i(\vec{1}) - Y_i(\vec{0})).
\end{equation}
Again, with random throttling the effect of interest is
$\tau_a^* = E(\tau_a)$, taking the expectation under the dropping
scheme $p(W)$. Note that $\tau_a$ considers the effect of treating all
eligible advertisers on the revenue generated {\em only} by advertiser
$a$, rather than the effect of treating only the queries for
advertiser $a$.

\section{RANDOMIZATION}
\label{section:randomization-estimators}
With {\em query randomization}, each query is randomized to treatment or control
and then all the eligible advertisers for the query are assigned to control if the query is
assigned to control or to treatment if the query is assigned to
treatment. Formally, an auction experiment is query-randomized if
\begin{enumerate}
	\item $P(Z_i = 1) = p_i$ and $P(Z_i = 0) = 1-p_i$, and
	\item $Z_i = 1$ implies that $\vec{Z}_{q[i]} = \vec{1}_{q[i]}$.
\end{enumerate}

With {\em (query, advertiser)} randomization, each (query, 
advertiser) pair is randomized independently to treatment or
control. Auctions now may have a mix of treated and control
bidders, which is not representative of the behavior of future
auctions. However, (query, advertiser) randomization may be necessary
if only some advertisers are allowed to be in the experiment, and
hence many of the auctions with treated advertisers will also have
untreated advertisers. Both query randomization and (query, 
advertiser) randomization happen before any form of throttling
takes place.

Our parameters of interest are total
differences over a day, not a mean per-query difference. 
Here the total under treatment is estimated by inversely weighting
each observation in treatment by its probability of occurrence, and
the total under the control is estimated by inversely weighting each
observation in control by its probability of occurrence.  

\begin{equation}\label{eq:tau-estimator}
	\hat{\tau}(Z) = 
\left 
  ( \sum_{i: Z_i = 1} \frac{Y_i(Z)}{p_i} - \sum_{i: Z_i = 0}
    \frac{Y_i(Z)}{1-p_i}  
\right)
\end{equation}

\begin{equation}
  \hat{\tau}_a(Z) = \left ( \sum_{i: a[i] = a, Z_i = 1} \frac{Y_i(Z)}{p_i} - 
  \sum_{i: a[i] = a, Z_i = 0} \frac{Y_i(Z)}{1-p_i}  \right).
\end{equation}

The estimators $\hat\tau$ and $\hat\tau_a$ would be unbiased for
$\tau$ and $\tau_a$ respectively, if the SUTVA assumption
(\ref{sutva}) held for payments $Y(Z)$, but SUTVA does not hold for
payments. Nonetheless, Section~\ref{section:theorems} shows
that $\hat\tau$ and $\hat\tau_a$ are unbiased under query
randomization.

\section{TWO EXAMPLES OF BIAS}
\label{models-bias}

To illustrate the issues, two toy examples show that the estimated
treatment effect for an experiment with just one auction can be
severely biased.

\subsection{Identical but independent bidders}

Suppose $K$ bidders participate in a first price auction, and they
all have the same bid under treatment and the same bid under
control. That is,
\begin{flalign*}
	B_i(1) &= R_1, \,\,\, i=1\ldots K\\
	B_i(0) &= R_0, \,\,\, i=1\ldots K\\
	R_1 &> R_0 .
\end{flalign*}

Also suppose each advertiser was independently assigned to treatment
with probability $\frac{1}{2}$; this is an example of (query,
advertiser) randomization. The goal is to estimate the effect of
treatment on this auction alone:
$\tau = \sum_{i=1}^K(Y_i(\vec{1}) - Y_i(\vec{0}))$. How ties are
decided does not matter. The expected value of the estimator defined
in \eqref{eq:tau-estimator} is
\begin{flalign*}
	E(\hat{\tau}) &= 
	\frac{1}{2^K} (\hat{\tau}(\vec{1}) + \hat{\tau}(\vec{0})) + 
\frac{1}{2^K} \sum_{Z \neq \{\vec{1},\vec{0}\}} \hat{\tau}(Z)\\
	&= \frac{1}{2^K}(R_1 - R_0) + \frac{1}{2^K} \sum_{Z \neq \{\vec{1},\vec{0}\}} R_1 \\
          &=\frac{1}{2^K} \tau + \frac{1}{2^K} (2^K - 2) R_1 \\
        &= \frac{1}{2^K} \tau+ (1 - \frac{1}{2^{K-1}}) R_1 
\end{flalign*}
where all $2^K$ possible assignments are equiprobable. The bias of $\hat\tau$ is
\begin{equation} \label{scenario1-bias}
	\mbox{Bias}(\hat{\tau}, \tau) = 
(\frac{1}{2^K} - 1) \tau + (1 - \frac{1}{2^{K-1}}) R_1.
\end{equation}
Intuitively, what happens is that the estimator $\hat{\tau}$ takes the
value $R_1$ for every assignment $Z$, except for $Z=\vec{0}$ where it
takes the value $-R_0$. Thus, as the number $K$ of bidders grows,
the bias approaches $R_0$, so the bias becomes as large as the control
bid as the size of the auction grows. For example, with $K=4$ bidders,
treatment bids of $R_1 = 6$ and control bids of $R_0 = 5$, the true
effect is $\tau = 1$ while equation~(\ref{scenario1-bias}) shows that
the bias is about 4.3. The same result would hold for second price
auctions in this scenario. The bias would not vanish if we only
allowed randomizations with given numbers $N_0$ and $N_1$ of control
and treated bidders respectively. Indeed, for any such
randomization scheme with $0 < N_0, N_1 < K$, the bias would be
exactly $R_0$ for any $K$. However, as will be shown in
Section~\ref{section:theorems}, the estimator is unbiased under query
randomization.

\subsection{Treatment  dominates control}

Suppose $K$ bidders participate in an auction and each
bidder under treatment bids more than every bidder under
control. If we label the bidders according to their bids under
treatment, then
\begin{flalign*}
	B_i(1) &> B_j(0) \,\,\, \mbox{ for all } i, j\\
	B_i(1) &> B_j(1) \mbox{ if } i < j
\end{flalign*}

Then (see the supplementary material) the bias of the estimated treatment
effect is bounded below by $\tau( \frac{1}{2^K} - 1) + A_K$, where
$A_K$ approaches $B_K(1)$ as the number $K$ of eligible bidders grows. Thus, 
the bias grows at least as large as $\max_{i}(B_i(0)) - (B_1(1) -
B_K(1))$.

The limiting bias is especially large if the smallest and largest bids
under treatment are
close, even if $K$ is small. When $K=4$, $(B_1(0), B_2(0), B_3(0),
B_4(0)) = (4, 4.25, 4.50, 4.75)$ and $(B_1(1), B_2(1), B_3(1), B_4(1))
= (6, 5.50, 5.25, 5)$, the true effect is $\tau = 1.25$, and the
exact bias is $3.8$, which is about three times as large as the effect itself.

\section{BIAS WITH  QUOTA THROTTLING}
\label{section:theorems}

\subsection{Joint and split  throttling}

Let $N_q[a]$ be the total number of queries that advertiser $a$ is
eligible for and let $Q[a]$ be the {\em quota} or number of queries
that advertiser $a$ can process. If $N_q[a] > Q[a]$, then the queries
for advertiser $a$ must be throttled so some are randomly dropped. An
experiment can use advertiser $a$'s quota for both treatment and
control, or it can split the quota into two pieces, using one piece to
service the $a$'s queries assigned to treatment and the other piece
to service $a$'s queries assigned to control. We do not
consider mixed experiments in which some advertisers are assigned to joint
throttling and others to split throttling.
\begin{definition} Under {\em joint throttling},
	\begin{equation}
		\sum_{i: a[i] = a} W_i(Z) \leq Q[a] 
	\end{equation}
	for all assignments $Z$ to treatment and all advertisers $a$. 
\end{definition}

\begin{definition} Under {\em split throttling},
	\begin{description}
		\item[] $\sum_{i: a[i] = a} I(Z_i = 1)W_i(Z) \leq Q^{(1)}[a]$,
		\item[] $\sum_{i: a[i] = a} I(Z_i = 0)W_i(Z) \leq
                  Q^{(0)}[a]$, and
		\item[] $Q^{(1)}[a] + Q^{(0)}[a] = Q[a]$
	\end{description}
	for all assignment $Z$ to treatment and every advertiser $a$, where
	$Q^{(1)}[a]$ and $Q^{(0)}[a]$ denote the {\em quotas} in treatment
	and control respectively.
\end{definition}

\subsection{Bid treatments and joint throttling}
\label{section:bid-level}

Suppose that the ad exchange always fulfills
each advertiser's quota entirely and that the number of queries that could
be sent to advertiser $a$ is more than it can process, so $N_q[a] >
Q_q[a]$. Then joint throttling implies that
\begin{equation}\label{eq:saturated-quota}
	\sum_{i: a[i] = a} W_i(Z) = Q[a] 
\end{equation}
for all advertisers $a$ and all assignments $Z$ to treatment. Further
suppose that throttling is random, so an advertiser is randomly
dropped from queries to meet its quota constraints. Then for all
vectors $w$ satisfying \eqref{eq:saturated-quota},
\begin{equation}
	P(W(Z) = w | Z) = \binom{N[a]}{Q[a]}^{-1}  \,\,\, \mbox{for all } Z.
\end{equation}
Under these assumptions, the following theorem is proved in the
supplementary material.
\begin{theorem} \label{bid-level-th}
With joint throttling and query randomization, the estimator
$\hat{\tau}$ is unbiased for $\tau^*$ for any bid treatment when every
advertiser is quota constrained and there are sufficient queries for
every advertiser's budget quota to be fulfilled.
\end{theorem}

Unsurprisingly, the same result holds for query randomization when 
all auctions are unconstrained, that is, when no advertiser is 
quota throttled. The corresponding theorem is stated and proved in the
supplementary material. In summary, query randomization leads to unbiased
estimates for bid treatments both with and without throttling.

\subsection{Quota treatment and split throttling}
\label{section:quota-level}
A quota experiment is designed to test whether a change to the
algorithm for dropping advertisers from queries to satisfy quota
constraints affects revenue. Suppose that the change is in addition to
the standard throttling algorithm (the control case) and that it is
applied before the standard throttling algorithm is applied. For
example, some feature of the query, such as the user's locale, could be
used to drop queries, reducing the need to randomly drop queries
without regard to their value to the advertiser. To be specific, the
treatment drops the (query, eligible advertiser) pair $i$ before the
control throttling algorithm is applied when a binary
random variable $x[i]$ is zero and the control
throttling algorithm alone is applied if $x[i]$ is one. Then the treatment
throttling enforces the constraint:

\begin{equation}
	(Z_i = 1) \cap (x_i = 0) \Rightarrow W_i(Z) = 0.
\end{equation}

The simulations in Section 6 show that the estimated effect of the
quota treatment on revenue is biased under joint throttling for both
query randomized and (query, advertiser) randomized experiments.  The
question is what happens if separate budgets are maintained for
treatment and control for each advertiser (i.e., under split throttling)?
Theorem~\ref{quota-level-th}, which is proved in the supplementary
material, shows that the estimated effect can be unbiased under
split-quota throttling under a set of conditions. Unfortunately, the
conditions are unlikely to hold in practice.

To state Theorem~\ref{quota-level-th}, let $N_a^{(1)}(x=1)(Z)$ be the
number of eligible queries
under treatment for advertiser $a$ when $x = 1$, and $Z = 1$. Define
$N_a^{(0)}(x=1)(Z)$ analogously. Also, define $N(x=1)$ to be the
total number of (query, eligible advertiser) pairs for which $x=1$.
\begin{theorem} \label{quota-level-th}
Let $\mathcal{Z}$ be the assignments of (query, eligible advertiser) pairs 
allowed by query randomization. If
	\begin{description}
		\item[] $x_i = 0$ implies that the bid $B_i$ for advertiser $i$ is 0,
		\item[] $\frac{Q_a^{(0)}}{N_a^{(0)}(Z)} =
                  \frac{Q_a^{(0)} + Q_a^{(1)}}{N_a}$ for all advertisers
                  $a \in \mathcal{A}$ and assignments of (query,
                  bidder) pairs $Z \in \mathcal{Z}$ to treatment and
                  control, and
		\item[] $\frac{Q_a^{(1)}}{N_a^{(1)}(x=1)(Z)} = 
           \frac{Q_a^{(1)} + Q_a^{(0)}}{N_a(x=1)} $ for all $a\in
           \mathcal{B}$ and 
           $Z \in \mathcal{Z}$
	\end{description}
	then the estimator $\hat{\tau}$ is unbiased for $\tau^*$ under
        query randomization with split throttling.
\end{theorem}

\section{SIMULATIONS}
\label{section:simulations}
This section reports the results of simulating the bias and variance
of revenue estimators under quota throttling for query randomized
experiments and (query, advertiser) randomized experiments. The simulated
query randomized experiments are {\em balanced} in the sense that they have
the same number of queries in treatment and control. Similarly, the
simulated (query, advertiser) randomized experiments balance the number of
(query, advertiser) pairs assigned to treatment and control. Without such
balance, effect estimates for total revenue would be much more
variable and that additional variability would obscure differences in
the randomization and quota sharing schemes for reasonably sized
simulations.

There are two main conclusions. First, when estimating {\em bid
  treatment} effects in constrained auctions, the estimated effects
under balanced query randomization are not only unbiased (as proved in
Theorem~5.1), but also have smaller variance than those obtained
with balanced (query, advertiser) randomization. Second, even if the
conditions of Theorem~5.2 are violated, query randomization combined
with split throttling can dramatically reduce the variance of the
estimated {\em quota treatment} effect, compared to the variance under
joint throttling.

Each simulated experiment considers auctions with three eligible advertisers. These
advertisers all compete in $N_q$ = 90, 120 or 150 auctions unless they are
throttled. Their quota limits are either $N_q / 3$ or
$2N_q/3$.  For bid treatments, potential bids are
$\mbox{Lognormal}(\mu_0 = 1, v=0.1)$ under control and
$\mbox{Lognormal}(\mu_1, v = 0.1)$ under treatment where $\mu_1$ is
either 1.05, 1.1, or 2. All the treatment bids in a simulation are
drawn from the same distribution, and all the control bids are drawn
from the same distribution. The control and treatment bids for a
(query, advertiser) pair are correlated, as described in the supplementary material. 
The potential bids define the true treatment effect on
total revenue.

For each distribution of potential bids and quota limit, we generated 20,000
random query experiments with exactly half the $N_q$ queries in each
experiment assigned to treatment and half to control. We also
generated 20,000 random (query, advertiser) pair experiments, with half the
pairs in treatment and half in control. The bid treatment effect or
quota treatment effect, depending on the nature of the simulation, was
estimated in each experiment. Because the bias depends on the
treatment effect for a lognormal, we report simulated bias relative to the true
effect.

\subsection{Bid treatments}
\begin{figure}[h!]
	\includegraphics[scale=0.4]{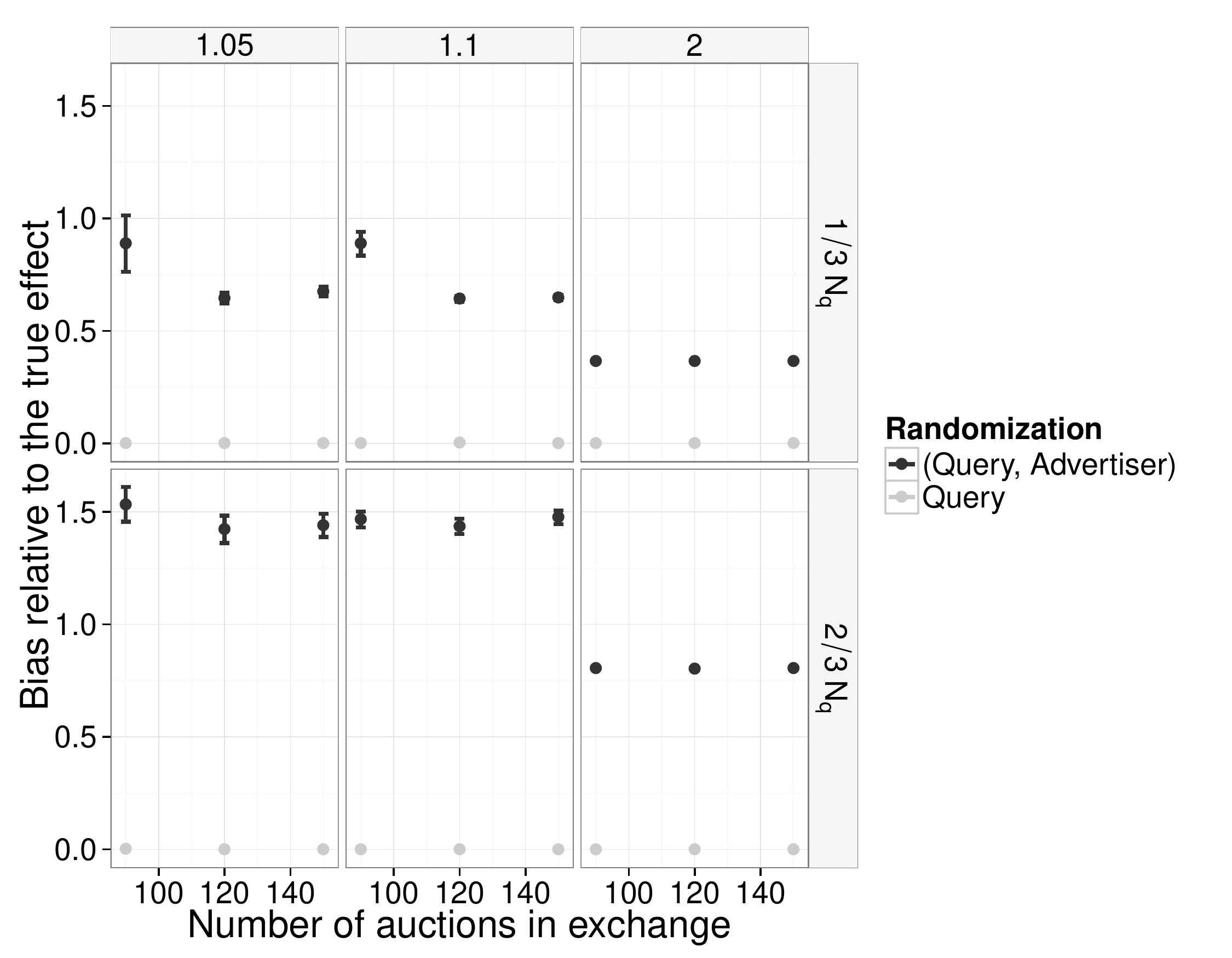}
	\caption{Simulated bias of the effect of a bid treatment
          relative to the true effect under balanced query
          randomization and balanced (query, advertiser)
          randomization. The dots show the mean relative bias, and the
          endpoints show the mean $\pm$ twice its simulated standard
          error. Columns show the mean log bid under treatment
        and the rows show the throttling rate.}

	\label{fig:bid_level_bias}
\end{figure}

Figure~\ref{fig:bid_level_bias} describes the simulated relative bias
(ratio of the bias to the true effect), where the rows correspond to
the different quota settings and the columns to the different mean log
treatment bids. The standard errors reported are computed over the 200
draws from the bid distribution and are divided by $\sqrt{200}$ to reflect
uncertainty about the simulated mean bias.

\begin{figure}[h!]
	\includegraphics[scale=0.38]{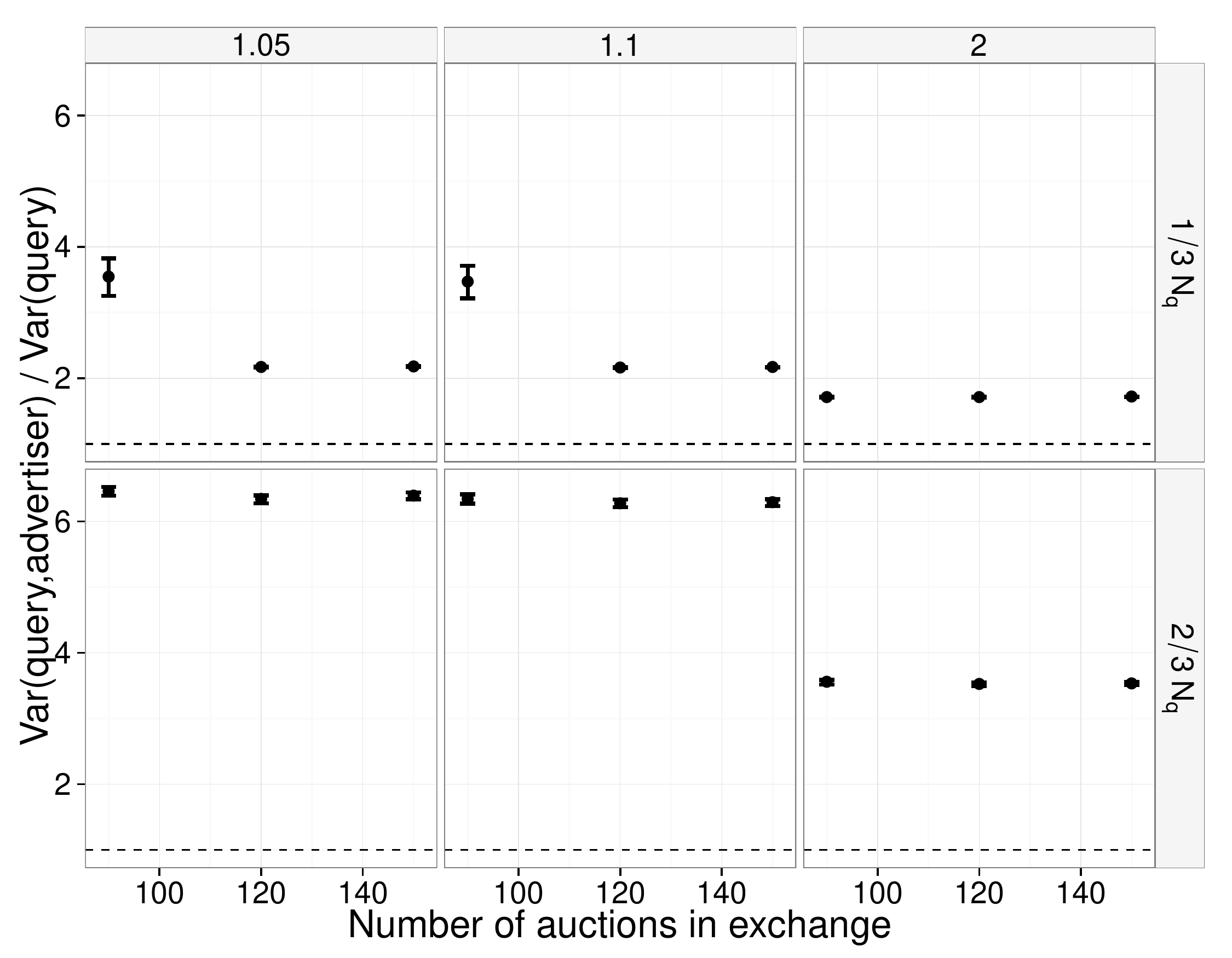}
	\caption{Ratio of the variance of estimated treatment effects
          under (query, advertiser) randomization to the variance of the
          estimated treatment effects under balanced query randomization for different lognormal
          bid distributions and throttling rates. The dots show the mean ratios
          and the endpoints show the mean $\pm$ twice its simulated
          standard error. The horizontal dotted line lies at one.}
	\label{fig:bid_level_var}
\end{figure}

Figure 1 shows that the relative bias is nearly zero for randomized
query experiments, while the relative bias has a mean as high as 1.5
for randomized (query, advertiser) pair experiments.  That is, the
penalty for allowing control and treated advertisers to compete in the
same auction is a 50\% increase in relative bias. Moreover, the
simulated variance of the bid treatment estimate under the random
query experiment is no more than its variance under the random (query,
advertiser) experiment (see Figure~\ref{fig:bid_level_var}.) The ratio of
simulated variances for (query, advertiser) randomization versus query
randomization is above one for all bid and throttling combinations
considered here, and as high as 6 when the treatment bids are 5\%
higher than the control bids, and the quota is around 66\%. Note that
the number of auctions in the experiment has little effect on the
relative bias or variance.

\subsection{Quota treatments}

\begin{figure}[h!]
	\includegraphics[scale=0.4]{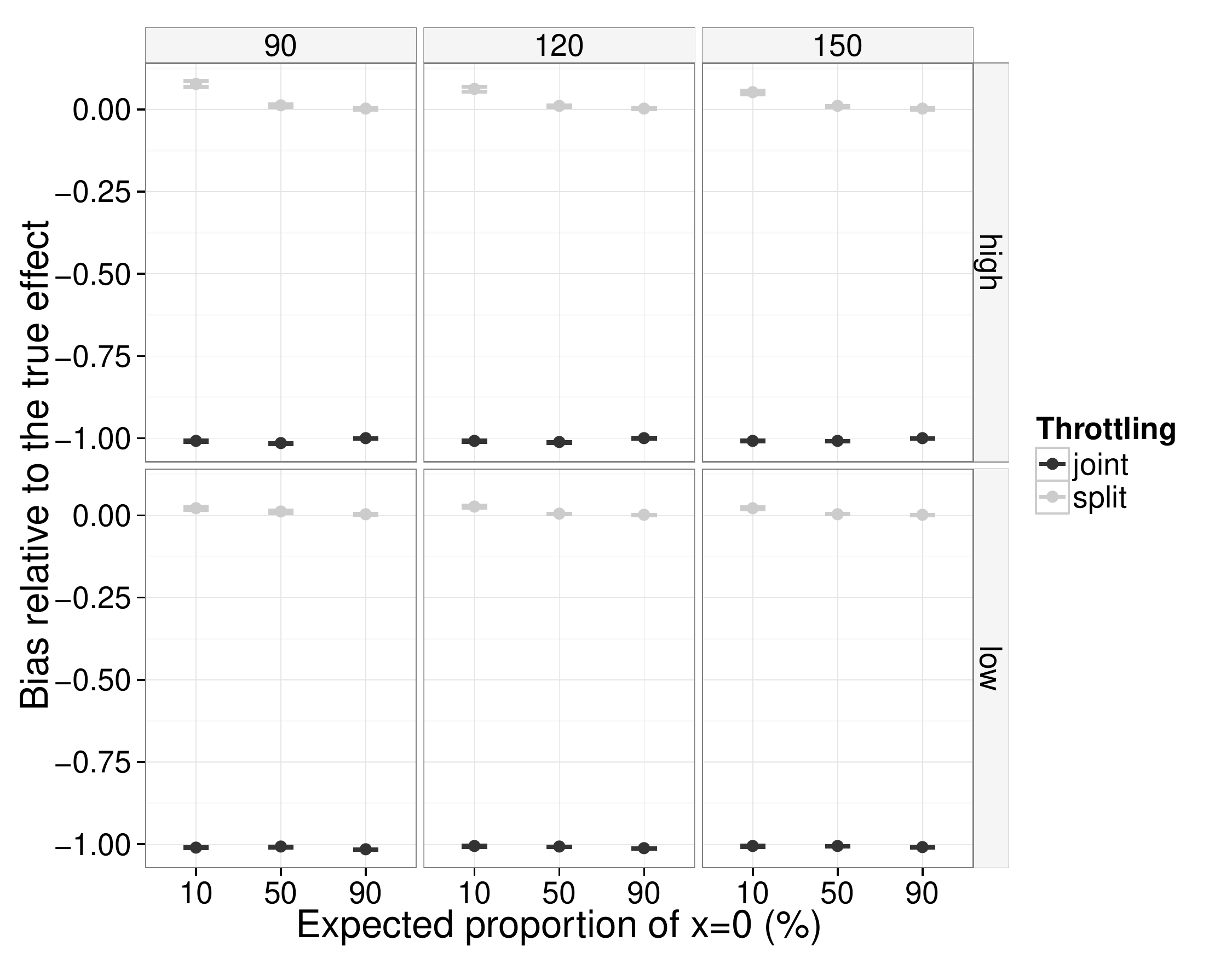}
	\caption{Bias of joint quota throttling and split quota
          throttling estimators relative to the true effect under
          query randomization. The dots are the mean relative bias
          and the segments show two simulated standard errors around
          that mean.}
	\label{fig:quota_level_bias}
\end{figure}

Because a quota treatment does not affect bids, the potential bids
with a quota treatment are the same under treatment and control. Here
$B_i(0) = B_i(1) \sim \mbox{Lognormal}(\mu_0 = 1, v=0.1)$. For each
random query or random (query, advertiser) pair, we draw a covariate
$x_i \sim Bernoulli(p_x)$, where $p_x = 0.1$ or $p_x = 0.5$ or
$p_x = 0.9$ in different
simulations. Figure~\ref{fig:quota_level_bias} shows the relative bias
of the estimate of total revenue under balanced query randomization
with joint throttling and with split throttling. The results for the
variance are shown in the supplementary material. Clearly, relative
bias is close to zero for split throttling (even if the conditions of
theorem~5.2 are violated), whereas it is around $-1$ for joint
throttling. This means that the estimator under joint throttling
estimates 0 regardless of the true effect on total revenue! So the
effect on total revenue of changing the throttling mechanism can never
be estimated from an experiment that uses joint throttling to meet
quota constraints. The supplementary material shows that this
conclusion about joint quota also holds for (query, advertiser)
randomization.

\subsection{Extension to larger samples}

Our simulations generate only a few auctions relative to the millions
of auctions that might participate in a real experiment every
day. Figures~\ref{fig:bid_level_bias} and \ref{fig:quota_level_bias}
hint that the relative biases are close to constant for larger numbers
of auctions, while Figure~\ref{fig:bid_level_var} hints that the ratio
of the variances is constant, or even slightly increasing as the
number of auctions in an experiment increases. Together, these figures
suggest that the variance of the estimator of the effect of a
treatment on total revenue increases faster under balanced (query,
advertiser) randomization than under balanced query randomization. We ran
additional simulations (see supplementary material) that give further
evidence for these trends.

\section{CONCLUSION}
Every day, companies like Google, Facebook and Yahoo run billions of auctions in ad
exchanges, and every day they experiment to improve the exchange. This
paper shows how casting experiments in the framework of potential
outcomes clarifies many practical issues, such as the consequences of
the choice of randomization. Bias has been emphasized throughout
because when it is large it makes experiments misleading.

As a general policy, query randomization should be preferred over
(query, advertiser) randomization when experimenting with bid treatments
because it allows an unbiased estimator of the true treatment effect,
without exceeding the variance of the same estimator under (query, advertiser) 
randomization. Split-quotas
are preferable to joint quotas when experimenting with throttling
treatments because split quotas have reduced bias and RMSE in simulations.

Admittedly, we do not have a complete solution to the problems that
arise in practice, such as the fact that advertisers are often
provided information about the user, such as location, that the
advertiser may use to decide whether to bid or the bid
amount. Experiments that take account of such covariates might be
better analyzed with statistical models rather than with the simple
estimators proposed here. Nor do we have analytical results for the
variance of the estimators or formal procedures for hypothesis
testing. These are all possible directions for future work.

\bibliographystyle{plainnat}
\bibliography{refs}

\end{document}